\documentclass[aps,prd,twocolumn,groupedaddress,nofootinbib,showkeys]%
{revtex4-1}

\usepackage{comment}

\usepackage[T1]{fontenc}
\usepackage{enumerate}
\usepackage{bm}
\usepackage{amsmath}
\usepackage{amssymb}
\usepackage{graphicx}

\newcommand{\scrif}{{\mathcal{I}^{+}}}
\newcommand{\scrip}{{\mathcal{I}^{-}}}

\newcommand{\const}{\mathrm{const}}
\newcommand{\conj}{\mathrm{conjugate}}

\begin{document}

\title{Fluxes and angular momentum}

\author{Adam D. Helfer}

\email[]{helfera@missouri.edu}
\affiliation{Department of Mathematics and Department of Physics \& Astronomy,
University of Missouri,
Columbia, MO 65211, U.S.A.}

\date{\today}

\begin{abstract}
The Ashtekar--Streubel fluxes give a proposed definition of the angular momentum emitted by an isolated gravitationally radiating system.  This was based on identifying a ``phase space of radiative modes,'' independent of any internal degrees of freedom, and using the Hamiltonian functions conjugate to the action of the Bondi--Metzner--Sachs (BMS) group as the energy--momentum, 
supermomentum and angular momentum.  
However, there are some difficulties in formulating this phase space so as to get the proper degrees of freedom.
I consider how to address this point, and also to identify circumstances in which the radiative modes are sufficiently decoupled that they can be assigned their own angular momentum.  Two different phase spaces are considered.  One, which seems to reflect what previous workers have done (it leads to the usual formulas for the Ashtekar--Streubel fluxes), is mathematically simpler, but this has unwanted degrees of freedom and is difficult to interpret physically.  The second is a quotient of the first; it is 
better justified (at least in terms of degrees of freedom), and plausibly decouples the radiative angular momentum from internal modes (at least for systems which ultimately become stationary).  The symplectic form for this quotient, and the corresponding angular momentum flux, involve highly non-local correlations.  Both phase spaces are shown to have Poisson brackets implementing the BMS algebra.
For both phase spaces, for axisymmetric space--times vacuum near null infinity, there can be no gravitational radiation of angular momentum about the axis of symmetry, although matter can carry off angular momentum in such cases.
\end{abstract}


\maketitle

\section{Introduction}

In an important paper, Ashtekar and Streubel \cite{AS1981} introduced the idea of defining a phase space of radiative modes of the gravitational field, and proposed using natural structures there to quantify the kinematics --- in particular, the energy--momentum and angular momentum --- of gravitational radiation.  The resulting formulas have been called fluxes (although they are in effect the time-integrals of what are usually called fluxes).

This work was a significant advance, for most previous attempts to deal with the question had had {\em ad hoc} features.  
The Ashtekar--Streubel approach, by contrast, was an attempt to build the analysis by appealing to general principles we may have some confidence in:  symplectic mechanics, and the relation between symmetries and conserved quantities.

Their starting-point was the introduction of a ``phase space of radiative modes'' $\Gamma_{\rm AS}$ of the gravitational field --- essentially, the space of allowable Bondi shears $\sigma$ at future null infinity $\scrif$.  There is a natural candidate $\omega_{\rm AS}$ for a symplectic form on $\Gamma_{\rm AS}$, given by formally taking the limit of the known symplectic form in terms of data on an acausal Cauchy surface.

Although no exact isometries can be expected to exist, there is what one might call a {\em weak symmetry group}, the {\em Bondi--Metzner--Sachs (BMS)} group of motions preserving the universal structure of $\scrif$.\footnote{The BMS group, although certainly expressing the invariance of the regime in question, is an infinite-dimensional object, introduced to compensate for lack of structure (or, equivalently, to take into account the breadth of allowable general-relativistic structures).  It is in this sense I call it weak.}   The BMS group will act on the space of Bondi shears, and Ashtekar and Streubel solved the equation
\begin{eqnarray} \label{eek}
  dH_\xi  =\omega_{\rm AS} (V_\xi ,\cdot)
\end{eqnarray}
to deduce the Hamiltonian functions $H_\xi$ conjugate to the Hamiltonian vector fields $V_\xi$ on $\Gamma_{\rm AS}$ associated with the BMS generators $\xi^a$.  It is these Hamiltonian functions which are called {\em fluxes}; they are given by Ashtekar and Streubel as certain integrals of the radiative data over $\scrif$.
Those conjugate to BMS translations they showed agreed with the Bondi--Sachs energy-momentum emitted in radiation; those conjugate to supertranslations gave the emitted Geroch supermomenta \cite{Geroch1977};  and those conjugate to BMS Lorentz motions they interpreted as the angular momentum emitted in radiation.\footnote{These classifications are relative to a given choice of Bondi retarded time parameter $u$.  Because of the supertranslation freedom in choosing this, in a BMS-based approach like this, there will be an infinite-dimensional family of angular momenta, indexed by the different cuts of $\scrif$.}

That this can be done in a relatively straightforward way is a consequence of a great deal of gauge control built into the structure.  The construction is phrased in terms of {\em characteristic data} (equivalent to the Bondi shear) for the radiation at $\scrif$; the gauge freedom is considerably reduced for such data.  

The proposal is natural, and is attractive in that it appeals to a structure (symplectic mechanics) we believe is deep, and does so in a way which is formally parallel to successful theories.  It is fair to say that this attractiveness is to some degree counterbalanced by the weakness of the BMS group --- it is not clear how this very large family of motions will produce as focussed a notion of conserved quantities as one has in non-general relativistic physics. But one would certainly like to push this analysis and see where it goes, taking as a guide the power of symplectic mechanics.

This program turns on the degree to which we can confidently identify the phase space.  It must encompass the relevant degrees of freedom; if there are any gauge degrees of freedom, they must be identified and controlled; finally, we must have good reason to think that these degrees of freedom can be isolated from others for the purposes of defining energy--momentum and angular momentum.

In this paper, I will examine the proposal in light of these concerns.
There are two, interrelated, issues.

The first is the question of the precise definition of the phase space, and in particular how it accommodates shears which do not vanish as the Bondi retarded time parameter $u\to\pm\infty$ and how it is acted on by supertranslations.  The difficulty here is that one does want to allow shears with non-trivial $u\to\pm\infty$ behavior, but at the same time one wants to carefully control any gauge freedom.  
This is a somewhat delicate degrees-of-freedom issue.
The Ashtekar--Streubel paper does not directly address this.

The second issue is to justify, for the purposes of computing the angular momentum, the assumed decoupling of the radiative modes.  This is not a simple thing.  Within the canonical framework, one would have to start from data for the entire system --- radiation and internal modes --- extend the BMS group action to at least the asymptotics of the internal modes, but then show show that, for at least a suitable class of systems, one could compute the kinematics (energy--momentum, angular momentum, supermomentum) of the emitted radiation without reference to the internal modes.  

The difficulty comes in splitting the physical degrees of freedom from the gauge modes.  This is delicate for angular momentum, because the degrees of freedom are typically functions of three real variables, but to account for the supertranslational freedom we must control the freedom even down to functions of two variables.
This issue does not appear explicitly in the Ashtekar--Streubel paper; it
is evidently bound up with degrees-of-freedom issue mentioned above.

As a particular example, suppose a system evolves to become significantly asymmetric, and then for an interval of retarded time emits gravitational waves, leaving a memory effect.  (This might happen, for instance, when a star becomes unstable, oscillates asymmetrically, and then supernovas.)  One physically plausible contribution to the emitted angular momentum would come from the ``cross product'' of the memory-induced supertranslation with the mass aspect.\footnote{The mass aspect is the quantity integrated over a cut of $\scrif$ to give the Bondi--Sachs energy--momentum.  The important thing in this example is that the the asymmetry means many multipoles may be present in the mass aspect, and these may pair against the multipoles in the supertranslation.}  This term would be {\em first-order} in the radiation data (and first-order in the matter, as reflected in the momentum aspect).  A term like this, coupling matter and radiation, if present, could well swamp the purely radiative effects.\footnote{Such terms do occur in the twistorial treatment of angular momentum \cite{ADH2007}. %
See also \cite{AB2017,AB2017b,BP2019,BGP2020} for investigations of possible such contributions from other definitions.}  
So we do need to know why, or in what circumstances, we may decouple the radiative angular momentum from other contributions.

One should also note that, to the extent the radiative angular momentum {\em does} decouple, it is difficult to assess its connection with other physics.  One really wants an accounting in which one can talk about the exchange of other forms of angular momentum with that in gravitational waves.  

Although the focus of this paper is on angular momentum, the identification of the radiative phase space is evidently of interest beyond that.  This space is used in an important way in Strominger's \cite{Strominger2014,SZ2014} argument that the Ashtekar--Streubel supermomenta are conserved and relate data on past and future null infinities.

Much of what is at issue in this paper depends on the $u\to +\infty$, $u\to-\infty$ behavior of the shear, and it will be helpful to have some terminology for this.  I will say a shear {\em has standard asymptotics} if it takes $u$-independent and purely electric\footnote{The ``electric'' and ``magnetic'' parts of the shear are conventional terms for parity-natural and parity-unnatural.  The precise rates of fall-off as $u\to\pm\infty$ will not be very important; for conceptual simplicity, I will assume that for $|u|$ large enough the limiting values are actually attained.}  values in the two regimes.  The values in the two regimes may be distinct; if they are, one (borrowing terminology from quantum field theory) says the shear {\em lies in an infrared sector}; otherwise it is in the {\em vacuum sector}.   Radiation memory is associated with shear in the infrared sectors.

When Ashtekar and Streubel define their phase space (see the paragraph in their paper containing their eq. (4.3)), they actually exclude infrared sectors:  the shear must fall off at least as fast as a power of $u$. 
This is presumably an oversight;
when they argue for the non-degeneracy of their symplectic form they do allow for the possibility of infrared-sector behavior.  One can also check that their calculations of their fluxes do not depend on being in the vacuum sector.  
For clarity, I will refer to the space of shears with standard asymptotics as the {\em extended phase space} $\Gamma_{\rm e}$.  This is presumably along the lines of what Ashtekar and Sreubel intended.

\subsection{Results}

The results are of two sorts.  First, there are conceptual clarifications, particularly of the possible structure of the phase space, but it should be made clear that conjectural elements remain.  Accepting these, however, the second class of results are flux formulas appropriate to shears which need not vanish as $u\to\pm\infty$ (in the case of the extended phase space, these just reproduce the Ashtekar--Streubel fluxes), and consequences of those.

\subsubsection{The phase spaces}

The main conceptual points go to the question of how to define the phase space so that infrared sectors are allowed but one is, as nearly as possible, isolating the physically relevant degrees of freedom and justifying, again as far as possible, the decoupling of the radiation's angular momentum from other contributions.  

We will see that there is a tension between using $\scrif$ as a hypersurface on which we collect the outgoing radiation profiles (its usual function) and wanting to use it for freely specifying data (its function in the flux approach to defining the phase space).  Also, even if we are ultimately interested in radiation, if from the start we {\em only} consider radiative data we have no way of taking up the question of whether its angular momentum decouples from other physics.

\begin{figure}[t!]
\centering
\includegraphics[trim= 7cm 14cm 0cm 5cm, clip=true,
scale=.2]{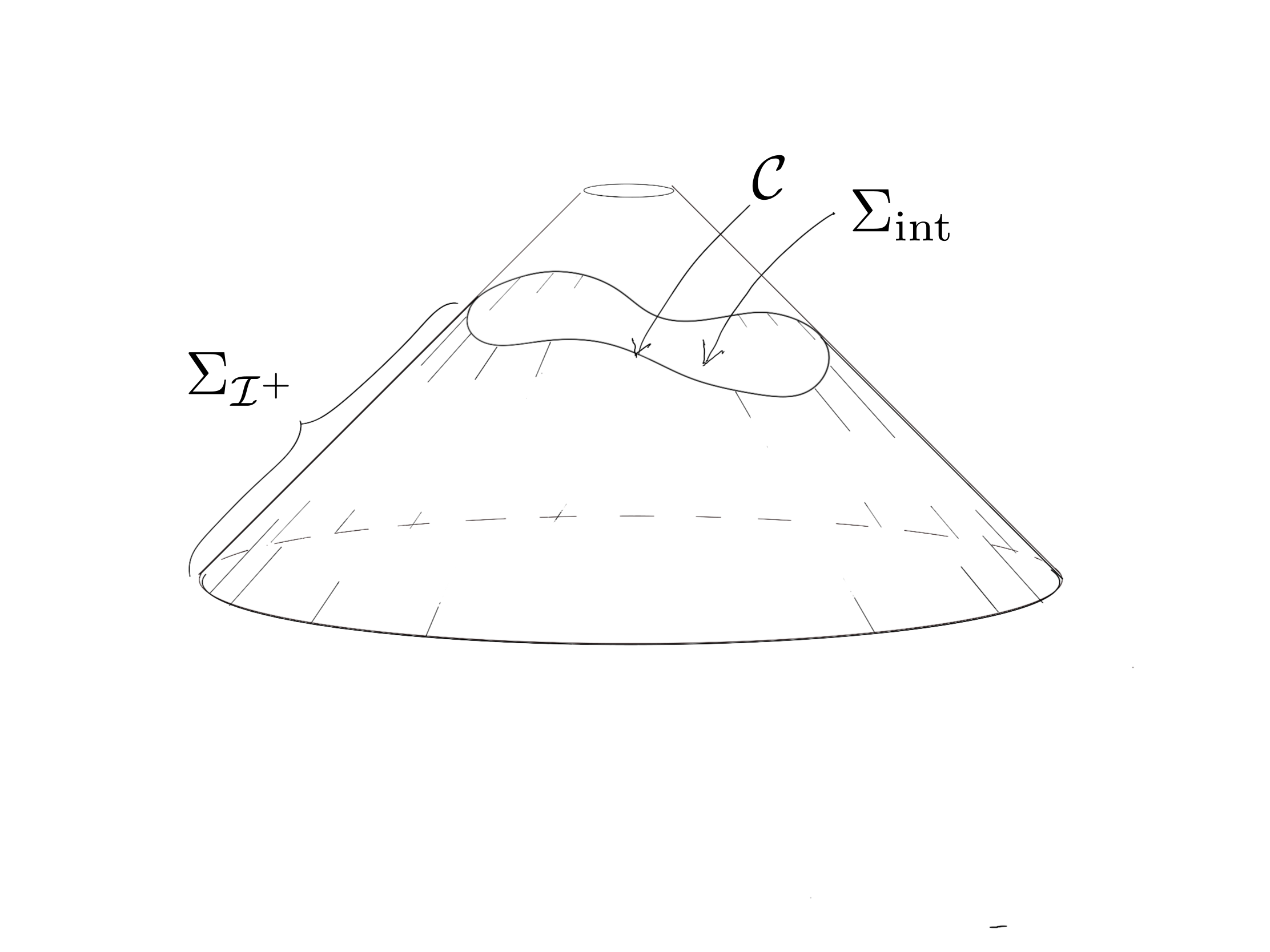}
\caption{A schematic representation of the Cauchy surface $\Sigma =\Sigma _\scrif\cup\Sigma _{\rm int}$.  The cone is future null infinity $\scrif$, and the space--time is below it.  The portion $\Sigma _{\rm int}$ contains interior data, and meets $\scrif$ at a cut ${\mathcal C}$; the portion $\Sigma _\scrif$ is the part of $\scrif$ to the past of that cut.
\label{fig:Cauchy}}
\end{figure}

It is possible to get a working compromise addressing these issues by appealing to considerations of well-posedness.  We want to begin by considering all possible degrees of freedom, but also to (as well as we can) split those into radiative ones and internal ones.  To do this, at least at a formal level, we may think of specifying data on a Cauchy surface $\Sigma$ which would contain, not all of $\scrif$, but the portion $\Sigma _\scrif$ of it prior to some late retarded time, together with an internal portion $\Sigma _{\rm int}$; see Fig.~\ref{fig:Cauchy}.  Physically, this corresponds to considering systems' radiation before the (late) retarded time.  Within this scheme, one can make a plausible case that the angular momentum of the radiation decouples, if the final system (represented by the data on $\Sigma _{\rm int}$) is stationary. 
Certainly, in many practical cases it is reasonable to suppose the final systems are stationary.  

As noted above, there are some choices to be made in the Ashtekar--Streubel paper in the precise definition of the phase space.  By the {\em extended phase space} $(\Gamma _{\rm e},\omega _{\rm e})$, I will mean the set of all shears with standard asymptotics, together with the Ashtekar--Streubel symplectic form on this space.\footnote{This has no obvious relation to the ``extended'' BMS algebra which has been considered by some authors (e.g.  ref.~\cite{FN2017}).}  It is this phase space which leads to the usual Ashtekar--Streubel fluxes.
We learn, though, from the considerations of well-posedness, that this space has unwanted gauge degrees of freedom --- for instance, it allows multiple representations of Minkowski space.  It may be possible to interpret these extra degrees of freedom as information needed to embed the isolated system represented by the Bondi--Sachs space--time in the larger Cosmos.

There is a formally natural way of defining a quotient $\Gamma$ of $\Gamma _{\rm e}$ which does not have the unwanted freedom, and there is a natural symplectic form $\omega$ on $\Gamma$, whose integral expression differs from $\omega _{\rm AS}$ by the addition of some interesting boundary terms at $u=\pm\infty$.  I do not have a proof that $(\Gamma ,\omega )$ is the ``correct'' phase space, but it is suggested by two lines of thought.

\subsubsection{The fluxes}

The BMS group acts naturally on each phase space $(\Gamma ,\omega )$, $(\Gamma _{\rm e},\omega _{\rm e})$, and one can work out the Hamiltonian functions conjugate to those motions --- the fluxes.  The results for the extended phase space agree with those of Ashtekar and Streubel.  
One striking feature is that the Ashtekar--Streubel fluxes are {\em inhomogeneous} in the shear, with the angular momenta and energy-momentum being purely second-order but the supermomenta having also first-order contributions.  
(In one natural way of writing the fluxes, these first-order contributions appear as 
``boundary'' terms associated with the limits $u\to\pm\infty$.) 
I will return to this discrepancy at the end of this subsection.

For the phase space $(\Gamma ,\omega )$, there are non-zero boundary contributions (in general) for the emitted angular momenta, but not for the emitted energy--momentum or emitted supermomenta.  
There is no homogeneity discrepancy, however:  all the fluxes are second-order in the shear.

An important point is that for each of the phase spaces $(\Gamma ,\omega )$, $(\Gamma _{\rm e},\omega _{\rm e})$ the Poisson brackets of the fluxes implement the BMS algebra.  The boundary contributions are necessary for this.  
This result is expected but also not quite trivial; it is certainly desirable, at least within the general framework of using the weak symmetry group as a basis for the program.

\begin{figure}[t!]
\centering
\includegraphics[trim= 1cm 2cm 0cm 1cm, clip=true,
scale=.225]{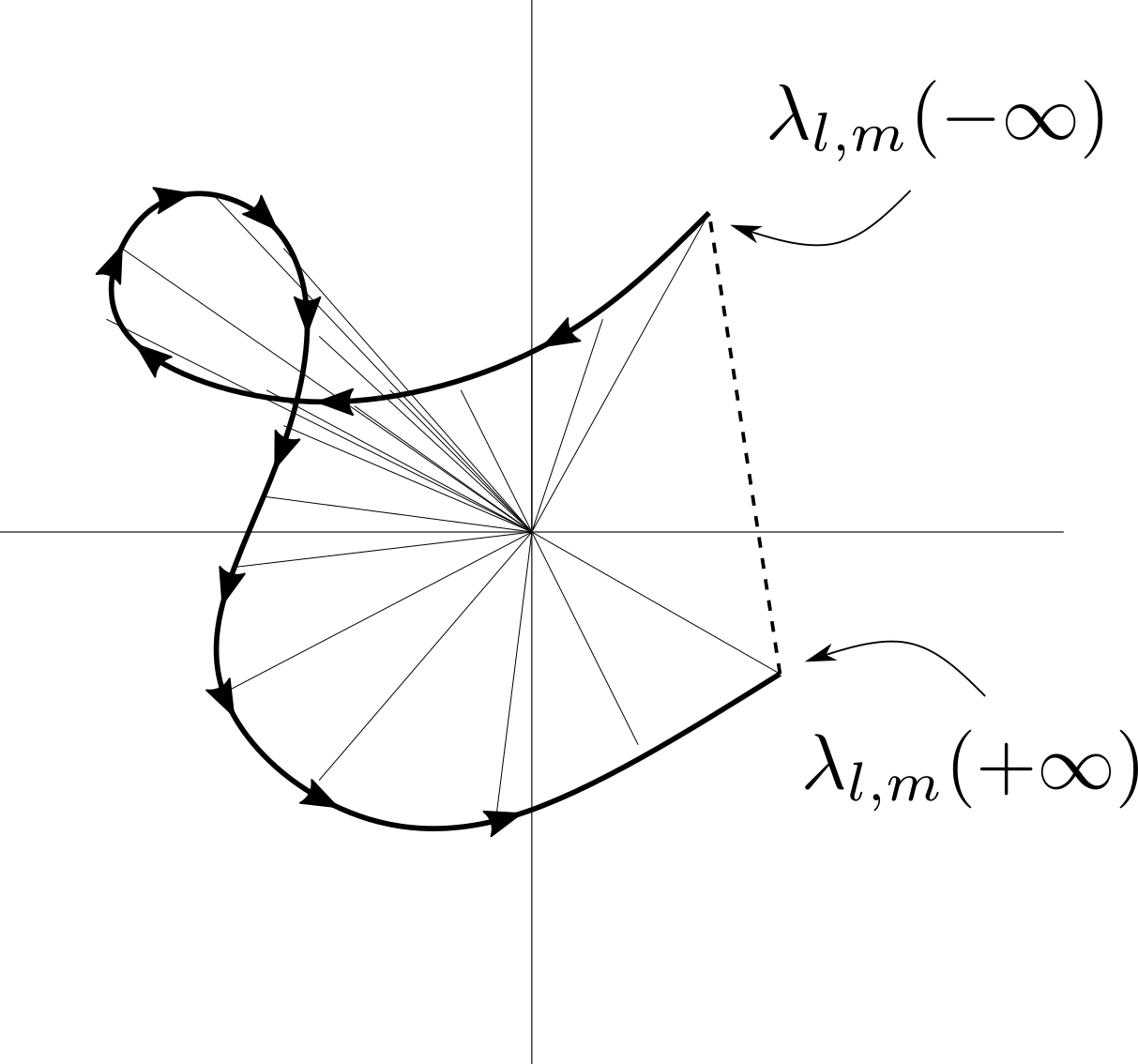}
\caption{The mode function $\lambda_{l,m}(u)$ determines a fan of radii from the origin in the complex plane, whose signed area weights the mode's contribution to the emitted extended angular momentum $H_{{\rm e},c}$.  The trajectory is the heavy line with arrows, and some of the radii of the fan are drawn.
For the angular momentum $H_c$ there is also a term from the straight-line trajectory from $\lambda_{l,m}(-\infty )$ to $\lambda _{l,m}(+\infty )$ (here shown dashed), counted with the opposite sign.
\label{fig:lambdaplane}}
\end{figure}

For spatial angular momentum about a given axis, I work out the fluxes in terms of a sum over spherical harmonic modes.  For each of these modes, the component of the shear can be thought of as a point $\lambda _{l,m}(u)$ in the complex plane, depending on $u$.  As $\lambda _{l,m}(u)$ moves, the radius from the plane's origin to the point will sweep out a fan.  The signed area of this, as $u$ ranges from $-\infty$ to $+\infty$ (times some factors) turns out to be the mode's contribution to the extended angular momentum (that is, for $(\Gamma_{\rm e},\omega _{\rm e})$), denoted $H_{{\rm e},c}$.  
This is a formally attractive result.
See Fig.~\ref{fig:lambdaplane}.

For the angular momentum $H_c$ for $(\Gamma ,\omega )$, there is an additional, boundary, contribution, which turns out to be {\em minus} the signed area one would have for a straight-line trajectory from $\lambda _{l,m}(-\infty)$ to $\lambda _{l,m}(+\infty )$.  
Alternatively, one can think of this as corresponding to the signed area of the fan swept out by the {\em closed} trajectory got by completing $\lambda _{l,m}(u)$ with the line segment.
This is an odd result, and one would like to know whether it is a hint of interesting unsuspected structure or a signal of pathology.

I also work out corresponding results for the Hamiltonian function conjugate to a boost, the center of energy.  (The expression for the decomposition of the shear in these modes appears to be new.)  The results are parallel to those for spatial angular momentum, taking into account the modes in this case form a continuum.

As I pointed out above, for the extended phase space there is a 
discrepancy in the order of the shear in the supermomentum vs. the other fluxes.
The reader may note that this is odd, since
in a BMS-based approach like this, one can define the angular momentum using any cut as the origin, and the change in passing from one cut to another brings in the supermomentum.  The discrepancy means there are no first-order terms when the cut is in the Bondi system, but angular momenta about supertranslated cuts will have 
first-order
contributions.
This raises questions about the naturality of the angular momentum fluxes.

Part of the resolution is that --- as the previous discussion indicated ---
the extended phase space contains non-dynamical modes.  In order to understand its significance in the case of weak dynamics (what we would ordinarily think of as weak gravitational waves), we should consider shears $\sigma = \sigma _0+\sigma _1$, where $\sigma _0$ is non-dynamical (purely electric and $u$-independent), and $\sigma _1$ is dynamical but uniformly small.  For such shears, the extended angular momentum has terms which are {\em first-order} in $\sigma _1$ (and first-order in $\sigma _0$).
This is important, because first-order terms have the potential to dominate the second-order ones.
However, as the scheme (as so far developed) is silent about how to assign or interpret $\sigma_0$, this represents an uncomfortable, potentially dominant, uncontrolled contribution
to the extended angular momentum.

\subsubsection{Consequences}

At the moment, it is not possible to argue definitively for (or against) any of these proposed formulas.  Each has what can be considered to be attractive features from some perspectives, but also also properties which are odd enough
to raise concerns.   
I already noted that the boundary contribution to the angular momentum $H_c$ was curious, and that interpretational issues need to be resolved for a secure application of $H_{{\rm e},c}$.
Beyond these,
there are some consequences of these results worth noting.

Although in one sense the modification of the phase space to $(\Gamma ,\omega )$ is quite minor, in another it is effectively a far less localizable concept (even at $\scrif$) than in the Ashtekar--Streubel treatment.  This is because the boundary terms are not simply a difference of some expression between $u=+\infty$ and $u=-\infty$, but involve products correlating the behaviors in those limits.  

Consider a system which is stationary except between two intervals $I_1$, $I_2$ of Bondi retarded time; these intervals themselves may be very greatly separated in retarded time.  One would like to suppose that one could model the changes of angular momentum due to each interval of emission by such a scheme, in other words, assuming that {\em each} of $I_1$, $I_2$ could ``effectively'' be considered to have a retarded time parameter $u$ running fro $-\infty$ to $+\infty$, and that the total angular momentum emitted would be the sum of these two contributions.  However, the boundary terms in general preclude this from holding.  (It does hold for momenta or supermomenta, but not for angular momentum.)

It is not clear what to make of this.  On one hand, in a practical sense one would indeed hope to be able to sum up the contributions to the angular momentum from these different intervals --- this is the principle behind the concept of flux (in the standard usage of the word).  On the other, the canonical or symplectic approach depends very strongly on considering a system as a whole; one cannot generally split it into subsystems without careful justification.  In this connection, it is important to remember that $\scrif$ is a {\em null} hypersurface, and so points on it can influence others --- although in a formal sense the {\em radiation data} can be freely specified on $\scrif$, one cannot regard the {\em physics} on one part of $\scrif$ as truly independent of the physics on another, causally related, part.  It may be that the non-additivity is reflecting this.

But all of the previous paragraph is to say that there are points on both sides and we await a better understanding of this.

Finally, both definitions of angular momentum have a remarkable property:  for an axisymmetric space--time (but without any other symmetry hypotheses), vacuum near $\scrif$, there will be no emission of angular momentum about the axis of symmetry.  This is curious, for it is easy to devise examples where material null radiation does lead to emission of angular momentum.\footnote{Consider a spinning ring, which at some time emits null radiation along its tangential directions, axisymmetrically.}  Why should it be so easy (in axially symmetric space--times) for matter to carry off angular momentum, but strictly forbidden for gravitational radiation to do so?  

It is not hard to see that, if we accept the expression $\int T_{ab}\xi ^a\, d\Sigma ^b$ for the angular momentum (where $T_{ab}$ is the stress--energy and now $\Sigma$ is a spacelike hypersurface extending to a cut of $\scrif$), we are led directly to the conclusion that only matter can carry it off (in the axially symmetric case).  In this sense the result is quite general.

It is worth contrasting this with the situation for timelike Killing vectors.  There, the corresponding conserved quantity is the energy, and in a stationary space--time neither material nor gravitational energy can be emitted.

\subsection{Outline}

Section II reviews the elements of the Bondi structure which will be relevant.

Section III clarifies the class of space--times we may expect the analysis to apply to, by applying considerations of well-posedness.  
Several conceptual points are uncovered, but the most important for the overall program is that even in the most favorable circumstances one must consider some coupling between the outgoing radiation and the final state of the internal space--time, and this coupling brings out issues of gauge freedom which were present but not apparent in the original work. 
Building on this discussion, the definitions of the phase spaces $(\Gamma ,\omega )$ and $(\Gamma _{\rm e},\omega _{\rm e})$ are given.

Section IV computes the Hamiltonian functions, and establishes some of their properties, in particular, that the Poisson brackets implement the BMS algera.

Section V gives the expressions for the spatial angular momentum and center of energy in terms of the modes of the rotation and boost operators, respectively.

The final section is given to discussion.

\subsection{Notation, conventions and background}  

While Ashtekar and Streubel worked in Geroch's \cite{Geroch1977} formalism, it is more convenient here to use that of Newman and Penrose \cite{NP1961}; see Penrose and Rindler \cite{PR1986} for this.  The papers of Dray and Streubel \cite{DS1984} and Dray \cite{Dray1985} provide useful formulas for comparing the formalisms.  The speed of light, and Newton's constant, are taken to be unity.

Many of the expressions in this paper include several terms and finish with 
the phrase ``$+\,\conj$.''  Unless this appears within parentheses, it is always to be understood as applying to all the preceding terms of the member.  Integrals over $\scrif$ are with respect to the standard volume form in a Bondi chart there; integrals over the sphere are denoted $\oint$ and are with respect to the standard area form.
The symbols $[\![\cdots]\!]$, $\langle\cdots\rangle$, giving the difference and average between quantities in the two regimes $u\to\pm\infty$, are defined in eqs. (\ref{jump}), (\ref{aver}).

\section{Preliminaries}

I will review here the key properties which will be used in this paper.  This section is not meant to be an exposition of the technical details (for which, see Penrose and Rindler \cite{PR1986}), but to give the reader an overall picture of the main structural features of null infinity and related concepts which will be important.

Throughout this paper, I will be concerned only with space--times admitting Bondi--Sachs asymptotics.  For such space--times, future null infinity $\scrif$ is a null hypersurface which has naturally the structure of a bundle of affine real lines over the two-sphere $S^2$.  It is conventionally parameterized by $(u,\theta ,\varphi )$, where $u$ is a {\em Bondi retarded time parameter}, and of course $(\theta ,\varphi )$ are polar coordinates on the sphere.  For Minkowski space, we can obtain $\scrif$ as a limit $r\to\infty$ of $x^a=ut^a +rl^a$, where $t^a$ is a unit future-directed timelike vector and $l^a$ is a null future-directed vector in the $(\theta ,\varphi )$ direction, normalized by $l_a t^a=1$, and $r$ is the usual radial coordinate.

It is convenient to use the spin-coefficient formalism \cite{NP1961,PR1986}.  In this, the two-dimensionality of $S^2$ and its orthocomplement are used to reduce tensorial quantities to components which have {\em spin weight}  (values in tensor products of the holomorphic and antiholomorphic line bundles).  The antiholomorphic angular derivative $\eth$ and its conjugate $\eth'$ play key roles.  

The BMS group acts on $\scrif$, respecting the fibration.  It is the semidirect product of the proper orthochronous Lorentz group with the {\em supertranslations}, which are the maps $(u,\theta ,\varphi)\mapsto \left( u+\alpha(\theta ,\varphi),\theta ,\varphi \right)$, with $\alpha$ a smooth real-valued function on the sphere.  A supertranslation is a translation iff $\eth^2\alpha =0$.  

A key quantity is the {\em Bondi shear} $\sigma (u,\theta ,\varphi )$, which has units length.  The {\em Bondi news} is $-\dot{\overline\sigma}$ (an overdot indicates differentiation with respect to $u$); the news is a measure of the gravitational radiation.  We can always write $\sigma =\eth^2\lambda$ for some spin-weight zero function $\lambda$, but in general $\lambda$ must be taken to be complex.  (And $\lambda$ will be unique up to a complex translation.)  The {\em electric} and {\em magnetic} parts of the shear are $\sigma _{\rm el}=\eth^2\Re\lambda$ and $\sigma _{\rm mag}=i\eth^2\lambda$.  (These terms are traditional; perhaps a better expression is that these are the natural-parity and unnatural-parity parts of the shear.)  

Under a passive BMS supertranslation the shear changes by $\sigma\mapsto \sigma-\eth ^2\alpha$, so the electric part of the shear changes but not the magnetic.\footnote{So, strictly speaking, the Bondi shear should be regarded, not as an ordinary function on $\scrif$, but as a functional also depending on the choice of Bondi coordinatization.}  In a non-radiating regime, the electric part of the shear can be gauged away by a supertranslation.  While magnetic shear is expected to be present generically, it is not clear that a $u$-independent purely magnetic shear can persist indefinitely, and the possibility is usually considered exotic.  

For these reasons, a system which radiates only within a bounded (but perhaps very large) interval of Bondi retarded time is generally modeled as having purely electric, and $u$-independent, shears in the regimes $u\to\pm\infty$.  Exactly how fast the shear should become $u$-independent, and how uniformly this should apply, are technical issues.  For simplicity, in this paper I am going to say {\em a shear has standard asymptotics} if for $|u|$ sufficiently large it is purely electric and $\dot\sigma =0$.\footnote{One can check that all that will actually by used in the computations is that $\sigma$ approaches $u$-independent electric limits as $u\to+\infty$ and $u\to-\infty$, and $\dot\sigma\to 0$ as $u\to\pm\infty$.}

Although in each of these regimes one could eliminate the shear by a supertranslation, there will usually be a gauge mismatch between the them.  This is one face of what is known as the {\em supertranslation problem}.  Physically, it means the two regimes cannot, even asymptotically, be related by a Poincar\'e motion.  This is a main source of difficulties in giving a good treatment of angular momentum at null infinity.  Roughly speaking, the shear (or more properly its potential $\lambda$) contributes a $(u,\theta ,\varphi)$-dependent problematic correction to the origin-dependence.

A cut with vanishing Bondi shear (in a Bondi coordinate system for which it is $u=$ constant) is called {\em good}, and a cut with non-zero shear is {\em bad}.  In a stationary Bondi--Sachs space--time, a Bondi coordinate system adapted to the stationarity will have vanishing Bondi shear.  

One point we shall have to consider is how much geometry at $\scrif$, beyond the shear, might enter into the physics under investigation.  For our purposes, it will be enough to bear in mind the following.

In measuring the energy--momentum at any cut, the {\em mass aspect}, denoted $\Psi _2^0+\sigma\dot{\overline\sigma}$ in the Newman--Penrose scheme, enters, where $\Psi _2^0$ is an asymptotic curvature coefficient.  For a knowledge of the emitted energy--momentum between two cuts, however, one does not need $\Psi _2^0$, but only $|\dot\sigma|^2$.  For the angular momentum at a cut, all workers would agree the curvature coefficient $\Psi _1^0$ should enter; for the angular momentum emitted between two cuts, different proposed definitions require different knowledge.  In the Ashtekar--Streubel scheme, knowledge of the shear would be enough.

In a stationary regime, considerable simplifications occur if the stationarity is expressed at $\scrif$ by $\partial /\partial u$ in a Bondi system.  In particular, the mass aspect is constant over $\scrif$.  (Recall that a mass aspect with multipoles, coupled with a memory effect, could plausibly contribute to emitted angular momentum.)

\section{The phase space}

Ashtekar and Streubel aimed to define a phase space of radiative modes of the gravitational field.  They took these to be be modeled essentially by the Bondi shear, with a particular symplectic form $\omega _{\rm AS}$; this can be justified as a formal limit of what one has when one takes an acausal Cauchy surface to null infinity (and ignores internal contributions).  
While it is certainly expected that radiation is coded in the Bondi shear, it is less clear that such a complete decoupling of internal degrees of freedom from radiative is justified for the problems at hand.

I will here consider the construction of the phase space.  I will show that there is a class of space--times for which the idea of considering the shears as radiative modes, for analyzing emitted angular momentum, does seem quite plausible.  These are the space--times which eventually become stationary.   
Although this is a teleological requirement, that at first seems not a serious restriction, because in many cases of practical interest we do expect the final system to be stationary.  There is, however, a subtlety, the nonlocality mentioned in the introduction, which does make the teleology a practical matter, and I will return to this later.
But leaving that aside, we have a structure which largely supports the Ashtekar--Streubel idea of defining a phase space of radiative modes.

When we come to include shears with non-trivial (but purely electric) limits as $u\to\pm\infty$, there are also a gauge degrees of freedom which must be considered.  This occurs because the shear is not wholly independent of the internal modes of the space--time.  
I will argue below that the most natural way of taking this freedom into account leads, not to the extended phase space $(\Gamma _{\rm e},\omega _{\rm e})$ of all shears with standard asymptotics, but to the quotient $(\Gamma ,\omega )$, and the fluxes for this quotient differ from those of Ashtekar and Streubel.

\subsection{Formal well-posedness}

To identify the degrees of freedom, we may begin with an initial-data surface.  (``Initial'' here is used just to indicate where the data are given.  It does not imply the surface lies in some distant past.)  To have a formally well-posed problem, it is not enough to have data on null infinity; we must make also some assumptions about the interior of space--time.

\subsubsection{Cauchy surfaces}

In the case of outgoing radiation, it is natural to take an M-shaped Cauchy surface $\Sigma =\Sigma _{\rm int}\cup \Sigma_{\scrif}$, where $\Sigma _{\rm int}$ is an achronal hypersurface in the interior but meeting $\scrif$ in a cut ${\mathcal C}$, and $\Sigma _{\scrif}$ is the portion of $\scrif$ to the {\em past} of ${\mathcal C}$; see Fig.~\ref{fig:Cauchy}.  Physically, this would correspond to looking at systems whose internal data were known at $\Sigma _{\rm int}$, and radiation emitted {\em prior} to ${\mathcal C}$.  
The portion of space--time to the future of $\Sigma _{\rm int}$ would be determined by the Cauchy data on $\Sigma _{\rm int}$, but the portion with the radiation we are tracking would 
have data on the whole of $\Sigma$ (so the Bondi shear on $\Sigma_\scrif$ would also be given) --- so $\Sigma$ would be a {\em final-data} (in a temporal sense) surface for this regime.  

It is clear that there is no objection, at a formal level, to using such a Cauchy surface.  There is nevertheless something very different here from one's usual view, in the interpretation given to the data on the different portions.  The issue is that
usually we do not think of specifying data {\em freely} at $\scrif$ --- far from it.  In most physical problems, we think of the degrees of freedom which will generate the radiation as specified on some internal hypersurface.  Future null infinity $\scrif$ is important, but as a place to record the outgoing wave profiles, not to adjust data at will.

It is worth noting that the time-reversed situation is more conventional.  
In this case, one has a W-shaped Cauchy surface $\Sigma ^{\rm past}_{\rm int}\cup\Sigma_{\scrip}$, with $\Sigma ^{\rm past}_{\rm int}$ an achronal internal hypersurface meeting $\scrip$ at a cut ${\mathcal C}_{\rm past}$, and $\Sigma _{\scrip}$ the portion of $\scrif$ to the {\em future} of ${\mathcal C}_{\rm past}$.  Here one thinks of internal data as given on $\Sigma ^{\rm past}_{\rm int}$, and data for radiation incoming thereafter freely specified.

In either one of these cases, the full phase space can be usefully thought of as having a sort of bundle structure, with the data at null infinity fibering over the space of data on the interior part of the Cauchy surface.\footnote{In principle, one could think of doing the fibration the other way, with base space the radiation data.  This seems not as natural, particularly if one considers the time-reversed case.}  (The radiation data must match up with the internal data at the relevant cuts.) 

There is a further issue to consider, which is the question of what restrictions we place on the data on $\Sigma _{\rm int}$.  If we want to think of the data on $\Sigma _\scrif$ as representing {\em all} of the outgoing radiation, then we want no radiation to arise to the future of $\Sigma_{\rm int}$, and the data there must meet this condition.  The simplest way to do this (and the only known way of ensuring it) is to require the space--time to be stationary to the future of $\Sigma _{\rm int}$; this also seems to be the most likely condition for decoupling the internal from the radiative angular momentum.\footnote{The argument for the decoupling is necessarily loose in the absence of a full understanding of angular momentum.  But the idea is that coupling is likely to occur through a mass aspect with high multipoles, but a stationary space--time will have only a monopole term in the frame determined by the stationarity.}

While this split into internal and radiative data is formally attractive, it does require knowing (or hypothesizing) the future evolution of the system.  We will look at consequences of this below.

\subsubsection{Gauge issues}

We now turn to the question of what sort of gauge issues there will be in this scheme, where stationary data are given on some internal hypersurface $\Sigma_{\rm int}$, and radiative data compatible with them are given on $\Sigma _\scrif$.  We will see that this leads to a difficulty due to the weakness of the BMS group. 

Let us break the problem down.  There will be two sorts of gauge motions:  those corresponding to deformations of $\Sigma _{\rm int}$ which fix its boundary ${\mathcal C}$, and those which alter the boundary.  The former are not troublesome; they correspond to evolution within the stationary portion of the space--time, do not bear on radiative effects, and are not of interest to us.

The difficulty comes because
we must also allow motions which give rise to supertranslations of ${\mathcal C}=\Sigma _{\rm int}\cap \Sigma _\scrif$.  This means that (even discounting the ``internal'' deformations of the previous paragraph), {\em the data specifying the space--time are not unique}.  For instance, Minkowski space would be represented by any purely electric $u$-independent shear.  More precisely, the data for $\Sigma _{\rm int}$ in this case would amount to embedding it into Minkowski space, with its boundary ${\mathcal C}$ some, generally bad, cut of $\scrif$, and the shear would be determined from that.  The same multiplicity-of-representation issue will apply to any of the space--times considered.
This is at odds with the basic idea of introducing a phase space which accurately codes the {\em physical} degrees of freedom.

The unwanted degrees of freedom are precisely the quotient Supertranslations/Translations, or equivalently the $u$-independent purely electric shears --- the weakness of the BMS group.  We must either alter the basic strategy of Ashtekar and Streubel, or find a way of eliminating this freedom.  We will consider both possibilities below.

This difficulty of getting the degrees of freedom correct leads to an awkwardness, or at least an unexpected behavior, which is worth pointing out.  Suppose we consider how, within this program, we assign fluxes to a particular space--time.  We imagine that we have the space--time, and we know in particular its shear.  What the argument above shows is that not quite all of the information in the shear should enter --- we should in some sense reduce that information by an action of Supertranslations/Translations, but there is no obvious candidate for such an action.\footnote{It may be worth emphasizing that we certainly do {\em not} want to reduce by anything beyond this quotient, since we wish to express the space--time's fluxes in terms of the $\scrif$ at hand.  
If we reduced by the Lorentz motions, we should not be able to assign {\em any} directional quantity, and so neither angular momentum nor center of energy.
If we reduced by the translations, it is immediately clear we would lose the center of energy; we would also lose the angular momentum, on account of its origin-dependence.}  We will see more specifically how this plays out as we pursue the program.

\subsection{The phase spaces}

The discussion above supports Ashtekar and Streubel's idea of using a suitable space of shears as a phase space of radiative modes, at least for space--times which become stationary for sufficiently late Bondi retarded time, provided we 
find some way of dealing with the unwanted freedom Supertranslations/Translations.

I will consider two possible resolutions of this.  One attempts to interpret the extra degrees of freedom; the other eliminates them.

\subsubsection{The extended phase space $(\Gamma _{\rm e},\omega _{\rm e})$}

Let $\Gamma _{\rm e}$ be the set of all shears with standard asymptotics; I will call this the {\em extended phase space}.  (It is most natural to take these shears to have conformal weight $-1$.)
The space $\Gamma _{\rm e}$ is attractive in its mathematical simplicity,
but we saw above that
it has unwanted degrees of freedom (for instance, all $u$-independent purely electric shears would specify Minkowski space).

The Ashtekar--Streubel symplectic form would be, 
on these shears and
in this notation,\footnote{I have adjusted the prefactor to give conventional normalizations for the fluxes.}
\begin{eqnarray}\label{oas}
  \omega _{\rm e}(s_1,s_2) =(8\pi )^{-1}\int_\scrif \{ s_1{\dot{\overline s}}_2 -
    {\dot s}_1{\overline s}_2\} +\conj\, .\qquad
\end{eqnarray}
Here $s_1$, $s_2$ are tangent vectors at some point in $\Gamma _{\rm e}$; these are functions on $\scrif$ which are thought of as infinitesimal perturbations of the shear.  
Ashtekar and Streubel show
that this is weakly non-degenerate.\footnote{Weak non-degeneracy means that $\omega_{\rm e} (s,\cdot )=0$, then $s=0$.  This is the appropriate non-degeneracy condition for a symplectic form in the present infinite-dimensional context.}

To suggest an interpretation of $\Gamma _{\rm e}$, let us note that Bondi--Sachs space--times are mathematical idealizations which are meant to model isolated systems within the Cosmos.  Any real isolated system will be modeled by a Bondi--Sachs space--time out to some finite but large regime ``near'' $\scrif$; beyond that, it must be matched to the the larger, ambient Cosmos.  Matching the metric will involve the full Bondi--Sachs data, not reduced by Supertranslations/Translations.  So it would seem reasonable to interpret the extra degrees of freedom of $\Gamma _{\rm e}$ as reflecting information needed to specify the embedding of the Bondi--Sachs space--time in the larger Cosmos.

This is only the germ of an idea, but it at least colorably suggests an interpretation of $(\Gamma _{\rm e},\omega _{\rm e})$.  Note, though, that with this view the extra degrees of freedom represent a coupling with the rest of the Cosmos, and so any fluxes we define may depend on this coupling as well as information intrinsic to the Bondi--Sachs space--time.

\subsubsection{The phase space $(\Gamma ,\omega )$}

I now come to the issue of trying to factor out the unwanted freedom from $\Gamma_{\rm e}$.  I do not have a first-principles resolution of this, but will suggest a solution which manifestly has the right degrees of freedom, comes independently out of two natural constructions, and is mathematically attractive.

I will first point out one possible approach, which I will reject.  
Recall that the extra degrees of freedom are precisely of the size $\text{Supertranslations/Translations}$.  One might try to find a BMS-induced action of this quotient group on $\Gamma _{\rm e}$, and reduce by that action.  Doing this would presumably mean finding a splitting
\begin{eqnarray}
  &&\text{Supertranslations} \\
  &&\qquad\cong \text{(Pure supertranslations)}\oplus \text{Translations}\, ;\nonumber
\end{eqnarray}  
one would project any supertranslation onto its ``pure'' (``translation-free'') part, and then use the BMS action of that.  There are several difficulties in doing this, however.  The most severe is that the symplectic form (\ref{oas}) fails badly to respect such a reduction, since one would have to allow $s_1$, $s_2$ to be {\em independently} acted on --- altered by {\em different} pure supertranslations.\footnote{Beyond that, making the split involves breaking Lorentz invariance (selecting a time-axis), and there is in this context no attractive way of doing that.  One could use value of the Bondi--Sachs energy--momentum at ${\mathcal C}$, but that would make the construction depend on more than the purely radiative data.  One could use the total energy--momentum emitted in gravitational waves, but this would make the phase space singular at the stationary space--times (since each of them has zero emitted energy--momentum, but is the limit of space--times with energy--momenta pointing in all different directions), and this would make the discussion of the weak-radiation-field limit problematic.}

It is better to consider physically how the extra degrees of freedom arose.  It was because we did not enforce matching conditions between the Bondi shear and data for the internal degrees of freedom.
We could eliminate the 
excess by considering, not the Bondi shear, but its $u$-derivative $\dot\sigma$ (equivalently, the news), as the quantity coding the gravitational radiation.  Then there are no matching conditions (we do, as before, require $\dot\sigma$ to vanish for sufficiently large $|u|$); the idea is to use whatever shear is induced by the {\em internal} data at the matching cut ${\mathcal C}$,
and find $\sigma$ elsewhere by integrating $\dot\sigma$ with that initial value.  This gets the degrees of freedom correct.  It is also conceptually attractive, in that one then has the data given cleanly as internal parts on $\Sigma _{\rm int}$ and radiation (just $\dot\sigma$, not $\sigma$) on $\Sigma_\scrif$.

It is convenient to represent this as a quotient $\Gamma$ of $\Gamma_{\rm e}$; we regard two elements of $\Gamma _{\rm e}$ as equivalent if they differ by a purely electric $u$-independent term.  In doing this, we are representing each $\dot\sigma$ by the equivalence class of $\sigma$ which give rise to it.  The freedom in this class is not one of supertranslations, but of ``constants'' of integration with respect to $u$, in recovering $\sigma$ from $\dot\sigma$.

At first, it appears this too will lead to a difficulty with the 
symplectic form, for $\omega _{\rm e}$ depends not only to the $u$-derivatives of the linearized shears, but on the linearized shears themselves. 
However, the issue here turns out to be much less problematic than for the attempt at reduction by Supertranslations/Translations.

Just how
confident should we be in the form $\omega _{\rm e}$?  
The argument for it is partly formal and partly rigorous.  One takes the conventional symplectic form for Einstein's theory, and deforms its initial-data hypersurface to $\scrif$.  The points of concern will be the control of the gauge, particularly at ${\mathcal C}$ and in the asymptotic regime encompassing $u\to -\infty$ on $\scrif$ and the asymptotics of the spatial hypersurface.  What I will show is that, while there is a strong sense in which the integrand of (\ref{oas}) does arise locally at $\scrif$ from this limiting procedure, one cannot discount the possibility that there are also boundary contributions, at $u=+\infty$ (our stand-in for ${\mathcal C}$) and $u=-\infty$.

If one {\em starts} with linearized perturbations in the Bondi--Sachs gauge (that is, perturbations which near $\scrif$ amount to changes in the non-universal fields in the Bondi--Sachs expansions), then there is no difficulty in deforming compact regions of the initial-data surface to $\scrif$ and the integrand (\ref{oas}) will arise as a matter of course.  Also, as Ashtekar and Streubel pointed out, Geroch and Xanthpoulos \cite{GX1978} showed that a broad class of linearized fields could, by change of gauge, be cast in that form.  But these arguments do not speak to what happens as we try to deform the {\em entire} initial surface to $\scrif$, and moreover the conventional symplectic form is itself not gauge-invariant unless restrictions are made on the linearized perturbations asymptotically; indeed, gauge changes give rise to boundary terms.

It therefore seems plausible to keep the integrand of eq. (\ref{oas}), but look for a ``boundary'' contribution to the symplectic form, involving the limits $u\to\pm\infty$, which will render it well-defined on the elements $\dot{s}$.

My propsal is to use
\begin{eqnarray}\label{symp}
\omega (s_1,s_2) &=&(8\pi)^{-1}\int _\scrif\{
   s_1\dot{\overline{s_2}} -{\dot s}_1\overline{s_2}\} \nonumber\\
   &&
     +(8\pi )^{-1}\oint_{S^2} \{ [\![ s_1]\!] \langle \overline{s_2}\rangle
       -\langle s_1\rangle [\![ \overline{s_2}]\!]\}\nonumber\\
       &&
       +\conj\, ,
\end{eqnarray}
where
\begin{eqnarray}
  [\![ \phi]\!] &=& \phi\Bigr| _{u=+\infty} -\phi \Bigr| _{u=-\infty}\, ,\label{jump}\\
  \langle \phi \rangle &=& (1/2) \left( \phi\Bigr| _{u=+\infty} +\phi\Bigr| _{u=-\infty}\right)\label{aver}
\end{eqnarray}
are the difference and average of the asymptotic values of a function of $\phi$ on $\scrif$ having well-defined limits, which may be functions of angle, as $u\to\pm\infty$.
It is straightforward to check that the formula (\ref{symp}) does respect the equivalence relation, and is weakly non-degenerate.
That it is BMS-invariant is shown by direct calculation, and the particulars of this will be discussed in the next section.

The boundary contribution in eq. (\ref{symp}) is really a sort of cross-term between the regimes $u\to\pm\infty$:  it can also be written
\begin{eqnarray}\label{altbound}
 && (8\pi)^{-1}\oint_{S^2}\left\{ s_1\Bigr| _{u=+\infty} {\overline s_2}\Bigr| _{u=-\infty}
    -s_1\Bigr| _{u=-\infty} {\overline s_2}\Bigr| _{u=+\infty}\right\} \nonumber\\
    &&\quad +\conj\, .
\end{eqnarray}    
For this reason, it cannot be realized via Stokes's theorem as the integral of the exterior derivative of some locally-defined expression in $s_1$ and $s_2$.

I noted above that there is a second way of arriving at the phase space $(\Gamma ,\omega )$.  We could have started with $\Gamma _{\rm e}$, and then tried to fix the gauge.  There are, however, more ways than one of doing this.  We might have required the shear to vanish as $u\to +\infty$, for example.\footnote{This would fit naturally with the discussion of well-posedness, and the idea of simplifying the data at ${\mathcal C}$.}  Doing this would have the effect of replacing the $s_j$ in the symplectic form by $s_j-s_j\Bigr| _{u=+\infty}$.  Alternatively, we could ask for the shears to vanish as $u\to-\infty$.\footnote{One motivation for this comes from the fact that the symplectic form $\omega _{\rm AS}$ is derived as a limit of integrals over acausal Cauchy surfaces.  To make such a limit rigorous, one must control the behavior of the integrands as the portions of these Cauchy surfaces near spatial infinity approach null infinity.  In such cases, tails of shears as $u\to -\infty$ could be problematic.}
In this case one would work with $s_j-s_j\Bigr|_{u=-\infty}$.

Actually, these two possibilities, and more, lead to the structure $(\Gamma ,\omega )$.  If we consider a weighted combination of these asymptotic conditions, say by working with
\begin{eqnarray}\label{aeq}
  {\hat s}_j=s_j -as_j\Bigr|_{u=+\infty} - (1-a)s_j\Bigr| _{u=-\infty}\, ,
\end{eqnarray}
for some real parameter $a$, then a short calculation shows
\begin {eqnarray}
  \omega _{\rm e}({\hat s}_1,{\hat s}_2) =\omega (s_1,s_2)\, ,
\end{eqnarray}    
independent of $a$.  So we see that any of the gauge-fixings effected by eq. (\ref{aeq}) would be equivalent to working with $(\Gamma ,\omega )$.

\section{BMS action and Hamiltonians}

At this point, the phase space $\Gamma$ has been defined as a space of shears (modulo a certain gauge freedom), equipped with a symplectic form $\omega$ (which is a ``conventional'' term plus some boundary contributions).  The action of the BMS group on the shears is well-known (and respects the gauge freedom).  We have therefore to invert the equation
\begin{eqnarray}
  dH_\xi = \omega (V_\xi,\cdot )
\end{eqnarray}
for the action $V_\xi$ of each BMS generator $\xi$ on $\Gamma$ to find the corresponding Hamiltonian function $H_\xi$.  This section does that, and verifies that the Poisson brackets implement the BMS algebra.

A few general points are worth recalling.  The phase space $\Gamma$ here is a vector space.  The condition that a linear map $s\mapsto Vs$ on this space infinitesimally preserve the symplectic form is $\omega (Vs_1,s_2)+\omega (s_1,Vs_2) =0$.  If this condition holds, the associated Hamiltonian function will be $H_V =(1/2)\omega (V\sigma ,\sigma )$.  If $W$ is also an infinitesimal symplectomorphism, then the Poisson bracket $\{ H_V,H_W\} = \nabla _{V\sigma}H_W =-\nabla_{W\sigma}H_V = -\omega (V\sigma ,W\sigma )$.

\subsection{Supertranslations}

The effect of an infinitesimal supertranslation by a function $\alpha$ on $S^2$ is to change the shear by $\alpha\dot\sigma-\eth^2\alpha$.  More formally, the vector field on $\Gamma$ will be
\begin{eqnarray}\label{suptrans}
  V(\alpha ) = \left(\alpha\dot\sigma -\eth ^2\alpha\right) \frac{\delta}{\delta\sigma}
    +\conj\, ,
\end{eqnarray}
where $\delta/\delta\sigma$ is a functional derivative, and on the right-hand side an integration over $\scrif$ is understood.

The vector field $V(\alpha )$ does not appear to be a linear function on $\Gamma$, because of the inhomogeneous term $-\eth^2\alpha$, but we must remember that $\Gamma$ is actually a quotient by such terms.  Therefore 
\begin{eqnarray}
\omega (V(\alpha ),s) &=&\omega (\alpha\dot\sigma -\eth^2\alpha ,s)\nonumber\\
  &=&\omega (\alpha\dot\sigma ,s)\, ,
\end{eqnarray}
and it is easy to see that $V(\alpha )$ is an infinitesimal symplectomorphism.  We will have
\begin{eqnarray}
  H_\alpha &=&(1/2)\omega (\alpha\dot\sigma ,\sigma )\nonumber\\
    &=& (16\pi )^{-1}\int_\scrif\{ \alpha |\dot\sigma|^2 -\alpha\ddot\sigma \overline\sigma\}
      +\conj\nonumber\\
    &=& (4\pi )^{-1}\int_\scrif\alpha |\dot\sigma|^2\, .
\end{eqnarray}
This is the emitted Geroch supermomentum.

Suppose on the other hand we had used instead the extended phase space.  In this case, we would have
\begin{eqnarray}
\omega _{\rm e}(V ,s)&=&\omega _{\rm e}(\alpha\dot\sigma -\eth ^2\alpha ,s)\nonumber\\
    &=&(8\pi )^{-1}\int_\scrif \{ (\alpha\dot\sigma -\eth ^2\alpha)\dot{\overline s} 
        -\alpha\ddot\sigma {\overline s} \}+\conj\nonumber\\
    &=&(4\pi )^{-1}\int_\scrif \alpha\dot\sigma\dot{\overline s} 
      -(8\pi )^{-1}\oint\eth^2\alpha[\![ \overline{s}]\!] \nonumber\\
      &&+\conj\, ,
\end{eqnarray}
which (one checks) would give the Hamiltonian function
\begin{eqnarray}
  H_{{\rm e},\alpha} &=& (4\pi )^{-1}\int_\scrif \alpha |\dot\sigma |^2\nonumber\\
  &&
    -(8\pi )^{-1}\oint \{ (\eth^2\alpha )[\![\overline\sigma ]\!]
      +({\eth'}^2\alpha)[\![ \sigma]\!]\}\, .\label{extsuperham}
\end{eqnarray}      
This is the Ashtekar--Streubel formula for the supermomentum flux.
The boundary term is first-order in the change in shear $[\![\sigma]\!]$.  It would vanish for momenta, but not generally for supermomenta.

\subsection{Lorentz motions}

A Lorentz motion is generated by a vector field $cm^a +{\overline c}{\overline m}^a$ on the sphere, where $c$ has spin-weight $-1$ and satisfies $\eth'c =0$.  I will denote the effect of this on the shear by
\begin{eqnarray}\label{DeltDef}
  \Delta \sigma = (c\eth +\overline{c}\eth' )\sigma  +k(-1+u\partial _u)\sigma
    +2(\eth c)\sigma\, ,
\end{eqnarray}
where
\begin{eqnarray}
  k=(1/2)(\eth c +\eth'\overline{c})\, .
\end{eqnarray}
(The BMS vector field is $\xi ^a = cm^a+{\overline c}{\overline m}^a +ku\partial _u$.  Because $\sigma$ is not really a scalar, but has spin-weight-two, the action of the angular terms on it is $c\eth\sigma +{\overline c}\eth'\sigma +2(\eth c)\sigma$.  Because $\sigma$ has conformal weight $-1$, one also gets a term $-k\sigma$ in eq. (\ref{DeltDef}).)

It is straightforward to verify that these BMS actions preserve both $\omega$ and $\omega _{\rm e}$.  (Equivalently, the three-surface integral terms and the boundary terms in the formula (\ref{symp}) for $\omega$ are separately preserved.)  It follows that corresponding Hamiltonians are

\begin{widetext}

\begin{eqnarray}
  H_c&=&(1/2)\omega (\Delta \sigma,\sigma )\nonumber\\
        &=& (8\pi)^{-1}\int_\scrif  (\Delta\sigma )\dot{\overline\sigma}
          -(16\pi )^{-1}\oint [(c\eth +\overline{c}\eth' -k+2(\eth c))\sigma]\overline\sigma
          \nonumber\\
          &&
          +(16\pi )^{-1} \oint \{ [\![ (c\eth +\overline{c}\eth' -k+2(\eth c))\sigma]\!]
            \langle\overline\sigma\rangle
            -\langle (c\eth +\overline{c}\eth' -k+2(\eth c))\sigma\rangle [\![ \overline\sigma ]\!]\}
            +\conj\nonumber\\
          &=&(8\pi)^{-1}\int_\scrif  (\Delta\sigma )\dot{\overline\sigma}
            -(8\pi )^{-1}\oint \langle(c\eth +\overline{c}\eth' -k+2(\eth c))\sigma\rangle [\![ \overline\sigma ]\!]\}
            +\conj \, ;\label{LorHam}\\
  H_{{\rm e},c} &=&(1/2)\omega _{\rm e}(\Delta\sigma ,\sigma )\nonumber\\
                  &=& (8\pi)^{-1}\int_\scrif  (\Delta\sigma )\dot{\overline\sigma}
          -(16\pi )^{-1}\oint [(c\eth +\overline{c}\eth' -k+2(\eth c))\sigma]\overline\sigma
          +\conj\nonumber\\
          &=& (8\pi)^{-1}\int_\scrif  (\Delta\sigma )\dot{\overline\sigma}
          -(16\pi )^{-1}\oint\{\langle (c\eth +\overline{c}\eth' -k+2(\eth c))\sigma\rangle [\![\overline\sigma ]\!] +[\![(c\eth +\overline{c}\eth' -k+2(\eth c))\sigma]\!]\langle\overline\sigma\rangle\}
         +\conj\nonumber \\
         &=& (8\pi)^{-1}\int_\scrif  (\Delta\sigma )\dot{\overline\sigma}
          +\conj
         \, .\label{LoreHam}
\end{eqnarray}         

\end{widetext}

\noindent (In passing to the last line, the boundary integrals cancel against their conjugates by integration by parts.)  Again, the formula $H_{{\rm e},c}$ is the Ashtekar--Streubel flux, in this case for angular momentum.

There are several points about these formulas worth noting:

(a) We may write
\begin{eqnarray}
 H_c&=&(8\pi )^{-1}\int_\scrif \left( \Delta\sigma -\langle\Delta\sigma\rangle\right)\dot{\overline\sigma} +\conj\, ,
\end{eqnarray}
where $\langle\Delta\sigma\rangle$ is the average of the values of $\Delta\sigma$ at $u\to+\infty$ and $u\to -\infty$.
This suggests that at least in some circumstances
the effect of the boundary term in eq. (\ref{LorHam}) is to cancel the preceding term.  As we will see below, this certainly need not always happen, but we note for now one case where the cancellation is exact:
Suppose $\sigma$ is a simple linear interpolation between its initial value $\sigma _-$ and its final value $\sigma_+$, say
\begin{eqnarray}
  \sigma =\begin{cases} \sigma _-&\text{for } u\leq u_-\\
         \frac{u_+ -u}{u_+-u_-}\sigma _- +\frac{u-u_-}{u_+-u_-}\sigma _+
              &\text{for } u_-\leq u\leq u_+\\
                 \sigma _{u_+} &\text{for } u\geq u_+\, .
                 \end{cases}\qquad
\end{eqnarray}               
Then it is easy to see that one does have $H_c=0$.

(b) The reader may notice something odd about the extended angular momentum:  the formula (\ref{LoreHam}) gives it as a purely degree-two function of the shear, but this refers only to the angular momentum relative to the family of cuts $u=\const$ in the chosen Bondi system.  In another, supertranslated, Bondi system, there would be supermomentum contributions, and these (according to eq. (\ref{extsuperham})) would be {\em first-degree} in the shear.  
This seems unphysical, in that the angular momentum should not depend so sensitively on the choice of cuts.

The issue here (which is bound up with the unresolved interpretational concerns about the extended phase space) is that there are non-dynamical degrees of freedom.  In particular, a supertranslation is not supposed to change the intrinsic physics, but it will change the shear inhomogeneously and therefore not preserve the degree of an expression in the shear.

We may get a more precise sense of what is going on by considering shears of the form $\sigma =\eth^2\lambda +\sigma _1$, where $\lambda$ is real and $u$-independent, and $\sigma_1$ is uniformly small; these may be considered to have first-order {\em dynamical} content $\sigma _1$.  Then we will have
\begin{eqnarray}
  H_{{\rm e},c}&=& (8\pi )^{-1}\oint (\Delta\eth^2\lambda) [\![\overline{\sigma _1}]\!]+\conj\\
  H_{{\rm e},\alpha } &=& -(8\pi )^{-1}\oint (\eth^2\alpha )[\![\overline{\sigma _1}]\!] +\conj\, ,
\end{eqnarray}
to first order in $\sigma _1$.  We see that in this case the angular momentum $H_{{\rm e},c}$ relative to the Bondi system, as well as the supermomentum, does have a first-order dynamical term, although those terms are only present if $[\![\sigma _1]\!]\not=0$.

The appearance of a contribution to the angular momentum which is first-degree in the dynamics is noteworthy.  On the other hand, that $H_{{\rm e},c}$ also depends on the non-dynamical $\eth^2\lambda$ underlines how important it is to sort out the interpretational issues.  Finally, I should comment that, although this first-degree behavior calls to mind the ``supernova'' example of the introduction, the two cases are significantly different.  The kind of term considered in the introduction was a sort of cross product between the change in shear $[\![\sigma]\!]$ and the mass aspect.  However, the mass aspect depends on more than just the radiation data, and so it cannot enter in the flux program.

(c) Let us consider space--times which are axisymmetric, with (the Bondi system and) $c$ chosen to give a rotation about the axis of symmetry.  For these, the angular momentum about the axis of rotation vanishes, for both the phase space $(\Gamma ,\omega )$ and the extended phase space $(\Gamma _{\rm e},\omega _{\rm e})$.  In other words, using either of these definitions, at least in the vacuum case, no angular momentum about the axis of symmetry can be radiated by gravitational waves.

In the case of the phase space $(\Gamma, \omega)$, since the curvature coefficient $\psi _3=-\eth\dot{\overline\sigma}$ will be axisymmetric, the news will be too (since $\eth$ is an injective rotation-preserving map from the spin-weight $-2$ to the spin-weight $-1$ functions).  We may then choose a representative shear to be axisymmetric, the condition for which is $c\eth\sigma +{\overline c}\eth'\sigma +2(\eth c)\sigma =0$.  This, together with $k=0$ (which holds for rotations) immediately leads to the vanishing of $H_c$.  

The argument for the extended phase space $(\Gamma_{\rm e},\omega _{\rm e})$ is a bit different.  In this case, while, as before, the news is axisymmetric, the shear could have a $u$-independent electric non-axisymmetric term. Only such a term could give a non-zero contribution to $H_{{\rm e},c}$ in eq. (\ref{LoreHam}), and such a contribution could come
only through the first factor in the integrand.  So we may replace $\Delta\sigma$ there by $\langle\Delta\sigma\rangle$, giving
\begin{eqnarray}
 H_{{\rm e},c}&=&(8\pi )^{-1}\oint \langle\Delta\sigma\rangle [\![\overline\sigma]\!]
   +\conj\, .
\end{eqnarray}   
However, an integration by parts converts this to
\begin{eqnarray}
 H_{{\rm e},c}&=&-(8\pi )^{-1}\oint \langle\sigma\rangle [\![\overline{(c\eth +{\overline c}\eth' -2(\eth' \overline{c}))\sigma}]\!]\nonumber\\
 &&
   +\conj\, ,
\end{eqnarray}   
and this vanishes, since only the axisymmetric part of the shear can contribute to $[\![\sigma]\!]$.

\subsection{Poisson brackets}

As noted in the beginning of this section, in the case that the phase space is a symplectic vector space and the infinitesimal symplectomorphisms are linear, the Poisson brackets of their Hamiltonian functions will implement the Lie algebra of the infinitesimal symplectomrophisms.  In our case, that means the BMS algebra is implemented by the Poisson brackets of the Hamiltonian functions $H_\alpha$, $H_c$ on $(\Gamma ,\omega )$; one can also verify this directly.

For the extended phase space $(\Gamma _{\rm e}, \omega _{\rm e})$, the supertranslations do not act linearly, and one needs a separate argument.  I will give one by explicit calculation.

It is easy to see that the brackets $\{ H_{{\rm e},\alpha _1},H_{{\rm e},\alpha _2}\}=0$ of the supertranslations among themselves vanish.  Also, although $\Gamma_{\rm e}$ should physically be considered a space with no preferred origin, it is mathematically a vector space, and the 
Lorentz motions {\em are} represented by linear maps on $\Gamma _{\rm e}$ and their Poisson brackets will implement the BMS Lorentz motions.  It remains only to check the Poisson bracket of a supertranslation and a Lorentz motion.

Let us denote by $\nabla _\alpha$ the derivative in $\Gamma _{\rm e}$ along the direction determined by the supertranslation (\ref{suptrans}) generated by $\alpha$.  Then we have
\begin{widetext}
\begin{eqnarray}
\nabla_\alpha H_{{\rm e},c} &=& (8\pi )^{-1}\int_\scrif
  \{ (\Delta (\alpha\dot\sigma-\eth^2\alpha)\dot{\overline\sigma } 
    +(\Delta\sigma )\alpha\ddot{\overline\sigma}\}+\conj\nonumber\\
    &=& (8\pi )^{-1}\int_\scrif
  \{ (\Delta (\alpha\dot\sigma-\eth^2\alpha)\dot{\overline\sigma } 
    -\alpha (\partial _u\Delta\sigma )\alpha\dot{\overline\sigma}\}+\conj\nonumber\\
    &=& (8\pi )^{-1}\int_\scrif \left( [\Delta ,\alpha\partial _u]\sigma)\right) \dot{\overline\sigma}
    -(8\pi)^{-1}\oint (\Delta\eth ^2\alpha)[\![ \overline\sigma]\!] 
        +\conj\nonumber\\
        &=& (8\pi )^{-1}\int_\scrif \left( ((c\eth\alpha +{\overline c}\eth\alpha) -\alpha k)\dot\sigma\right)  \dot{\overline\sigma}
    -(8\pi )^{-1}\oint (\Delta\eth ^2\alpha)[\![ \overline\sigma]\!] 
        +\conj\nonumber\\
        &=&H_{{\rm e},(c\eth\alpha +{\overline c}\eth\alpha - k\alpha)}\, ,
\end{eqnarray}        
\end{widetext}
where the last step requires a short spin-coefficient calculation
to verify $\eth^2\left( c\eth\alpha +{\overline c}\eth\alpha - k\alpha\right) 
= \Delta\eth^2\alpha$.  This is the correct Poisson bracket, and one should note that the boundary term for the extended-phase space supermomentum enters essentially; without it, the Poisson brackets would not implement the BMS algebra.

\section{Angular momentum}

The previous section derived the Hamiltonian functions conjugate to the motions induced by the BMS vector fields, for the two phase spaces $(\Gamma ,\omega )$, $(\Gamma _{\rm e},\omega _{\rm e})$.  In particular, those which are in the flux program interpreted as the relativistic angular momenta $H_c$, $H_{{\rm e},c}$ were found, and they differed by a boundary term; the boundary term had, roughly, the effect of canceling some of the other contributions to $H_c$.  

In this section, I examine the relativistic angular momentum in more detail, both in the case of (ordinary) spatial angular momentum, and the center of energy (more often called center of mass).  In each case, the
shear is decomposed into suitable modes, each one-complex-dimensional, and the angular momentum $H_c$ is shown to be equal to a sum (or integral) over these of what might be called the {\em phase area} swept out in each of these complex planes by the component of the system in that mode over
the course of the system's evolution.  The effect of the boundary term is to remove the phase area corresponding to a straight-line motion, from the initial to the final point, in each of these planes.

One consequence of this is that emission of angular momentum requires a change in the argument (angle in the complex plane) of at least some modes of the system.

\subsection{Spatial angular momentum}

For the case of spatial angular momentum we will choose $cm^a+{\overline c}{\overline m}^a$ to be the generator of rotations about (say) the $+z$ axis.  
We may write
\begin{eqnarray}
\sigma &=&\eth ^2\sum_{l,m} \lambda _{l,m} Y_{l,m}\, ,
\end{eqnarray}
where the coefficients $\lambda _{l,m}$ are complex-valued functions of $u$.  
Here only the values $l=2,3,\ldots$ will contribute, and $m$ ranges from $-l$ to $l$, as usual.
Then we have
\begin{widetext}
\begin{eqnarray}
 H_{{\rm e},c} &=& (8\pi )^{-1}\int_\scrif \left(\eth^2\sum_{l,m}  im\lambda _{l,m}Y_{l,m}\right)
  \left(\eth '^2\sum_{l,m}  \dot{\overline{\lambda _{l,m}}} \overline{Y_{l,m}}\right) +\conj
  \nonumber\\
  &=&(8\pi )^{-1}\sum_{l,m}  \oint(\eth^2Y_{l,m}{\eth'}^2\overline{Y_{l,m}})
    m\int\left(  i \lambda_{l,m}\dot{\overline{\lambda_{l,m}}}-i\overline{\lambda_{l,m}}{\dot\lambda}_{l,m}\right)\, du \nonumber\\
    &=&(8\pi )^{-1}\sum_{l,m}  (1/4)(l+2)(l+1)l(l-1)
    m\int\left(  i \lambda_{l,m}\dot{\overline{\lambda_{l,m}}}-i\overline{\lambda_{l,m}}{\dot\lambda}_{l,m}\right)\, du
    \label{spatang}
    \, .
\end{eqnarray}    
\end{widetext}
For each pair $(l,m)$ of mode indices, the integral is twice the signed area in the complex $\lambda _{l,m}$-plane swept out by the radii from the origin to $\lambda _{l,m}(u)$.  We might call this the {\em phase area} of the mode, for the change is the change in phase.
See Fig.~\ref{fig:lambdaplane}.

The contribution of the boundary term would evidently be minus what one would have by linearly interpolating between the end-points, which is to say minus the signed area of the triangle connecting the end-points.  
An equivalent statement is that the total contribution got by the signed area of the fan determined by first moving along the trajectory $\lambda _{l,m}(u)$, and then closing this by the segment from $\lambda _{l,m}(+\infty )$ back to $\lambda _{l,m}(-\infty)$.

\subsection{Center of energy}

It turns out that it is possible to get results for center of energy (sometimes called center of mass) parallel to those for the spatial angular momentum.  The analysis is a bit different, though, for two reasons.  First, we must take account of an explicit contribution from moments of the radiated energy--momentum, which is not present in the spatial case; and second, the 
eigenvalues of the generator of boosts form a continuum.

As before, I will first discuss the flux $H_{{\rm e},c}$ without the boundary terms.  Recall that this is given by
\begin{eqnarray}
  H_{{\rm e},c} = (8\pi )^{-1}\int _\scrif (\Delta\sigma )\dot{\overline\sigma} +\conj\, ,
\end{eqnarray}
where 
\begin{eqnarray}
\Delta\sigma =   (c\eth +\overline{c}\eth') \sigma +k(-1+u\partial _u)\sigma +2(\eth c)\sigma\, .
\end{eqnarray}
We will take $c=\eth' \cos\theta$ (in standard polar coordinates); this generates a unit boost in the $+z$ direction.  Then $k=-\cos\theta$.

The contribution to $H_{{\rm e},c}$ explicitly involving a moment of the radiated energy--momentum is
\begin{eqnarray}
H_{{\rm rad},c}=(4\pi )^{-1}\int_\scrif ku|\dot\sigma|^2\, .
\end{eqnarray}
It is always possible to remove this by a supertranslation.  In fact, in generic circumstances, one can do much more.  To see this, note that
for each generator of $\scrif$, there will be a unique real value $u_0=u_0(\theta ,\varphi )$ such that
\begin{eqnarray}
  \int (u-u_0)|\dot\sigma|^2\, du =0\, ,
\end{eqnarray}
as long as (what will be true generically ) $|\dot\sigma|^2$ is non-zero somewhere along the generator.  Moreover, this value of $u_0$ will be BMS-covariant, since along each generator $u$ changes only by an affine motion under a BMS transformation.  Therefore, as long as $|\dot\sigma|^2$ is non-zero somewhere along each generator, we can by a unique supertranslation arrange to have $u_0(\theta ,\varphi )=0$ and in particular $H_{{\rm rad},c}=0$.

To take care of the non-generic cases, note first that, if $\dot\sigma$ vanishes identically, the result is trivial.  Otherwise $k$ will be non-zero almost everywhere (in fact, except at $\theta =\pi /2$), and we can find some generator where $k\not=0$ and we have $\dot\sigma\not=0$ somewhere.  Then by choosing $u_0(\theta ,\varphi)$ to be a suitably scaled bump function in a neighborhood of this generator, we can arrange a supertranslation giving $H_{{\rm rad},c}$ vanishing.

We now assume that by a suitable supertranslation we have $H_{{\rm rad},c}=0$, so that
\begin{eqnarray}
  H_{{\rm e},c} &=& (8\pi )^{-1}\int _\scrif \left((c\eth +\overline{c}\eth') \sigma -k\sigma +2(\eth c)\sigma \right)\dot{\overline\sigma} \nonumber\\
  &&+\conj\, .
\end{eqnarray}
We now make the usual polar-coordinate choice of spin-frame.  The angular operator $D$ appearing in this integral given by
\begin{eqnarray}
D\sigma &=& (c\eth +\overline{c}\eth') \sigma -k\sigma +2(\eth c)\sigma \nonumber\\
 &=& -\sin\theta\partial_\theta \sigma -\cos\theta\sigma\nonumber\\
   &=& -\partial _\theta (\sin\theta \sigma) 
\end{eqnarray} 
is evidently anti-self-adjoint with respect to the usual Hermitian inner product (of spin-weight $2$ functions) on the sphere.  
A short calculation shows the eigenfunction with eigenvalue $i\mu$ is
\begin{eqnarray}
\varepsilon _\mu =\pi ^{-1/2} \csc\theta \left[\frac{1+\cos\theta}{1-\cos\theta}\right]^{i\mu /2}\, ,
\end{eqnarray}
and an arbitrary dependence on $\varphi$.  The normalization has been chosen so that we have the orthonormality condition:
\begin{eqnarray}
&&\int \varepsilon _{\mu}(\theta )\overline{\varepsilon _{\mu}(\acute\theta )}\, d\mu
       \nonumber\\
  &&\qquad = 2\csc\theta\csc\acute\theta \delta\left( \log\left[
    \frac{1+\cos\theta}{1-\cos\theta}\cdot
     \frac{1-\cos\acute\theta}{1+\cos\acute\theta}\right]\right)\nonumber\\
   &&\qquad = 2 \csc\theta\csc\acute\theta \delta (\cos\theta -\cos\acute\theta ) 
     (\sin^2\theta ) /2\nonumber\\
     &&\qquad = \delta (\cos\theta -\cos\acute\theta ) \, .
\end{eqnarray}
(In passing from the second to the third line, we use the identity $\delta (f(x)) =\delta (x-x_0)/|f'(x_0)|$ when $f$ has a single simple zero at $x_0$.)

The complementary relation 
\begin{eqnarray}
\int _0^\pi\overline{\varepsilon _{\acute\mu}}\varepsilon_{\mu}
      \sin\theta\, d\theta &=& \delta (\mu-\acute\mu)
\end{eqnarray}      
can be derived
by
integrating by parts:
\begin{eqnarray}
  &&\int _\eta^{\pi-\eta} \left( \overline{\varepsilon _{\acute\mu} }D\varepsilon_{\mu}
    -\varepsilon _\mu \overline{D\varepsilon_{\acute\mu}}\right) \sin\theta\, d\theta
    =
    -{\overline{\varepsilon _{\acute\mu}}}\varepsilon_\mu \sin ^2\theta \Bigr| _\eta^{\pi-\eta}\nonumber    \\
&&    i(\mu -\acute\mu) \int _\eta^{\pi-\eta}\overline{\varepsilon _{\acute\mu}}\varepsilon_{\mu}
      \sin\theta\, d\theta 
      =
      -\pi ^{-1}\left[\frac{1+\cos\theta}{1-\cos\theta}\right]^{
      i(\mu-\acute\mu )/2}\Bigr| _\eta^{\pi-\eta}\nonumber  \\
    && \int _\eta^{\pi-\eta}\overline{\varepsilon _{\acute\mu}}\varepsilon_{\mu}
      \sin\theta\, d\theta 
      =
       -\frac{2}{\pi} \frac{\sin \left(\frac{\mu-\acute\mu}{2}\log\frac{1-\cos\eta}{1+\cos\eta}\right)}{\mu
        -\acute\mu} \qquad\qquad 
\end{eqnarray}      
and taking the limit $\eta\downarrow 0$.

Thus any shear $\sigma$ can be written as
\begin{eqnarray}
  \sigma =\int \sigma_\mu   \varepsilon _{\mu} \, d\mu\, ,
\end{eqnarray}
for
\begin{eqnarray}
  \sigma_\mu  =\int _0^\pi \sigma \overline{\varepsilon _{\mu}}\, \sin\theta\, d\theta\, ,
\end{eqnarray}
where $\sigma_\mu$ will also generally depend on $u$ and $\varphi$.  
In terms of this, we have
\begin{eqnarray}\label{coe}
  H_{{\rm e},c}&=& (8\pi )^{-1}\int \left(\mu\int_\scrif \left( i\sigma_\mu \dot{\overline{\sigma_\mu }} -i\overline{\sigma_\mu}{\dot\sigma}_\mu\right)\right)\, d\mu
   \, .
\end{eqnarray}

The formula (\ref{coe}) just derived, for the ``extended'' center of energy, is a close parallel to the one (\ref{spatang}) for the ``extended'' spatial angular momentum, and parallel comments apply.  The inner integral of (\ref{coe}) gives twice what I called the phase area of the mode.  Also the same parallel applies to the difference between $H_{{\rm e},c}$ and $H_c$:  the contribution of the boundary term would be to subtract, mode for mode, the phase area got from the straight-line trajectory from the initial to the final point.

\section{Discussion}

Ashtekar and Streubel proposed to define the angular momentum emitted in gravitational radiation by identifying a phase space of radiative modes, and treating the BMS motions as symmetries which should have conjugate constants of motion --- those conjugate to Lorentz motions would be angular momenta.  This idea is formally attractive, and offers the opportunity to connect gravitational radiation with canonical mechanics.  At the same time, one should bear in mind that the BMS motions constitute only what I called a {\em weak} symmetry group, an infinite-dimensional one introduced to compensate for lack of structure.

There was a technical question, though, of just how to define the phase space; and also
whether one could justify the assumed decoupling of the radiative angular momentum from other details of the geometry of the system.  This paper has been concerned primarily with those issues.  (See refs. \cite{AB2017,AB2017b,BP2019,BGP2020} for other work related to the second issue.)

I was led to two candidate phase spaces $(\Gamma ,\omega )$, $(\Gamma _{\rm e},\omega _{\rm e})$, on each of which the BMS group acted, and on each of which fluxes could be defined whose Poisson brackets implemented the BMS algebra.  The 
extended fluxes are just those of Ashtekar and Streubel, but the fluxes for $(\Gamma ,\omega )$ differ, outside of the vacuum sector,
by certain boundary contributions (at $u=\pm\infty$).
There are unresolved questions about each.
I will sketch the arguments leading to these.

There was, to begin with, a tension between thinking of $\scrif$ as the hypersurface where outgoing wave profiles are tracked (its usual interpretation) and thinking of it as a place to freely specify data for gravitational processes (the idea behind a phase space of radiative modes) --- usually we do not think of the modes in the future as freely specifiable.  I suggested what appears to be a working compromise,
based on formal well-posedness, by using data on a Cauchy surface $\Sigma _\scrif\cup\Sigma _{\rm int}$, with $\Sigma _{\rm int}$ representing the final state of the system and $\Sigma _\scrif$ the portion of $\scrif$ on which the radiation is recorded.  I took the final state to be stationary, which both ensured that all the radiation had been accounted for on $\Sigma _\scrif$ and made the separation of radiative from other angular momenta plausible.

Taking into account the gauge freedom led to the phase space $(\Gamma ,\omega )$, where it was necessary to introduce a boundary term to the integral for the symplectic form in order to maintain the proper invariance.  This boundary term is arguably formally attractive, but it does have the effect of making the expression for $\omega$ far more non-local than $\omega _{\rm e}$.  

The assumption that the final state be stationary is teleological, but at first does not seem very problematic practically, since in practice many systems of interest will have stationary final states.  There is however a subtlety, which comes on account of the non-locality of the boundary terms in the symplectic form:  Suppose that a system radiates only in two intervals $I_1$, $I_2$ of Bondi retarded time, very separated from each other.  We would like to think that we could view each of these intervals as ``effectively'' extending from $-\infty$ to $+\infty$, and compute the total angular momentum emitted as the sum of the emissions from the two intervals.  But the boundary terms prevent us from thinking of the phase space $(\Gamma ,\omega )$ as a direct sum of phase spaces $(\Gamma _{I_1},\omega _{I_1})$, $(\Gamma _{I_2},\omega _{I_2})$, and of taking the total angular momentum emitted to be the sum of the emissions for the two intervals.  

So the issue for $(\Gamma ,\omega )$ is not just one of teleology, but of {\em comprehension:}  the construction does not give a well-defined emitted angular momentum unless we are sure that we have included {\em all} the radiative periods.  While arguably the structure of $(\Gamma ,\omega )$ is attractive, both mathematically and in bringing in an element of physical globality, the issue of comprehension does present a difficulty for its practical application.

I also considered the extended phase space $(\Gamma _{\rm e},\omega _{\rm e})$, which does not involve the same passage to a quotient, but (as a result) contains non-dynamical modes.  (For instance, Minkowski space had an infinity of representations.)  I suggested that these extra degrees of freedom may be interpretable as needed for specifying the isolated system's embedding within the Cosmos, but, until this is made precise, the structure is not fully controlled.  In particular, if we accept this interpretation, the angular momentum of the radiation cannot be wholly separated from the embedding.

The extended phase space does not suffer from the problem of comprehension.  Its angular momentum flux is significantly different from in that {\em the extended angular momentum flux} $H_{{\rm e},c}$ {\em may have contributions which are first-degree in the dynamical modes,} although these contributions are also proportional to non-dynamical terms.  Such terms could well dominate in many cases, and this underscores the need for an interpretation of the non-dynamical modes.

A formula for the spatial angular momentum in terms of the rotational modes (spin-weighted spherical harmonics) was worked out.  For the spatial angular momentum, the shear was a sum of the mode functions times
complex coefficients $\lambda _{l,m}(u)$, and the angular momentum was a sum of integrals determined by the trajectories $\lambda _{l,m}(u)$.  Curiously, the effect of the boundary terms
for $(\Gamma ,\omega )$ was to close each of these trajectories by a line segment from $\lambda _{l,m}(+\infty)$ to $\lambda _{l,m}(-\infty)$.  Parallel results hold for the center of energy.

An especially interesting question is whether axisymmetric space--times can radiate gravitational angular momentum. 
In special relativity, the angular momentum is given by $\int T_{ab}\xi ^a\, d\Sigma ^b$ (where $T_{ab}$ is the stress--energy and $\Sigma$ is a suitable spacelike hypersurface), and it has long been suggested that this formula can be taken over to general relativity, at least when $\xi ^a$ is genuinely a Killing vector.  If this {\em is} accepted, then all the angular momentum arises locally from the stress--energy, and in particular, if no stress--energy escapes to $\scrif$ there can be no purely gravitational radiation of angular momentum.  Both of the flux definitions considered here have this feature.

To appreciate the force of this, I will temporarily invoke some common but potentially problematic concepts, and consider that the gravitational field might be described by ``gravitons,'' and --- what will be the key point --- that these and other quanta can, to some degree, interconvert.  This is done loosely in many places, and indeed it is a generic feature of any naive attempt to quantize general relativity in parallel to other field theories (see e.g. \cite{BCDS2012} and references therein).  It is extensively considered in discussions of the Gerstenshtein process \cite{Gertsenshtein1962}, an interconversion of photons and gravitons in the presence of a classical electromagnetic field.
It is important to understand that, while this language is supposed to describe underlying microphysics, we are still assuming that, macroscopically, the situation can be modeled by classical relativity with a classical stress--energy.  

Consider a variant of the system described earlier.  As before, there will be an axisymmetric rotating body, which at some point emits, axisymmetrically and tangentially, some matter (``matter'' just means something carrying stress--energy), and that matter will partially convert into gravitons.  
The loss of the converted quanta will alter the angular momentum, but the gravitons, or more properly, the gravitational disturbance created, cannot compensate for this by carrying any of that angular momentum across $\scrif$.  It must remain in the combination of the modified field and the remaining matter.  There is necessarily a back-reaction, so that $\int T_{ab}\xi ^a\, d\Sigma ^b$ is preserved.

In particular, this means that, if matter does not escape to $\scrif$, there is some limit to its conversion to gravitons, for enough matter must remain to carry the angular momentum.

As I noted above, it is a bit dubious to speak of ``gravitons;'' the arguments here are really only suggestive.  But they are enough to make us ask what constraints conservation of angular momentum puts on matter in the axisymmetric case


%

\end{document}